\documentclass[12pt,aps]{revtex4}
\usepackage[dvips]{graphicx}
\newcommand{\be}{\begin{equation}}
\newcommand{\ee}{\end{equation}}
\newcommand{\bea}{\begin{eqnarray}}
\newcommand{\eea}{\end{eqnarray}}
\newcommand{\ci}{\cite}
\newcommand{\bi}{\bibitem}
\newcommand{\nono}{\nonumber \\}
\newcommand{\tr}{{\rm tr}}

\newcommand{\e}{{\rm e}}

\newcommand{\ssF}{{\sin^2 F}}

\newcommand{\da}{\dagger}
\newcommand{\dd}{\partial}
\newcommand{\bftau}{\mbox{\boldmath$\tau$}}

\newcommand{\half}{\frac{1}{2}}

\def\dal{\,\lower0.3ex\vbox{\hrule\hbox{\vrule\kern2pt\vbox{\kern4pt\kern4pt}
\kern2pt\vrule}\hrule}\,}

\def\s{\sigma}
\def\o{\omega}
\def\L{{\cal L}}

\begin{document}
\begin{center}
{\bfseries SKYRME  MODEL LANGUAGE \\
IN THE THEORY OF NUCLEONS AND NUCLEI }\\[15mm]

{\hfill\parbox{9cm}{\small\tt Talk at the International Workshop on Nuclear Theory, Rila, June 2001, to be published in the Proceedings.}}

\vskip 15mm
 Nikolaev V.A.$^{1}$,
Tkachev O.G.$^{2}$\\[5mm]
(1){\it
INRNE, Sofia , Bulgaria
}\\
(2){\it
Institute of Physics and Information Technologies,
Far East State University,\\ Vladivostok, Russia
}\\
{\it E-mail:

    $^{1}$nikolai@spnet.net \space
    $^{2}$tkachev@ifit.phys.dvgu.ru
}
\end{center}
\setcounter{footnote}{2}
\footnotetext{This work has been partially supported by grant "Universities of Russia" N 015.02.01.022}
\setcounter{footnote}{1}
\vskip 10mm

\begin{center}
{\bf Abstract}
\vskip 10mm

\begin{minipage}{15cm}
{\small
In this talk we try to clarify the problems existing on the way of theorist
decided to construct nuclear theory on the generalized Skyrme model
background. We conclude that to construct such a model of light nuclei
one have to construct a hybrid model where one particle degrees of freedom
are concentrated around the surface of the nuclei and  soliton with
non-trivial structure is located at the center region.
}
\end{minipage}
\end{center}

\vskip 10mm
\section{Introduction}
\indent

In this talk we try to clarify the problems existing on the way of theorist
decided to construct nuclear theory on the generalized Skyrme model background.
The Lagrangian appropriate for a generalized Skyrme model in
the leading classical field approximation
yields chiral soliton solutions.  This solitons are associated with
the nucleon or multibaryons  and will be used for obtaining spatial structure information
on baryons or baryon systems.

The present paper includes discussion of our recent work, extensions,
and calculations of the nucleon electromagnetic form factors in the
generalized Skyrme model as well. This model involves the theoretical
description of the dilaton-quarkonium scalar field and shows its
importance in the description of soliton dynamics.  We use
"dilaton-quarkonium" scalar field to indicate the way we
subdivide the gluon condensate in the calculation.  This generalized
model reproduces the experimental value of the nucleon mass, the input
being the experimental value of the pion decay constant and the
theoretically derived value of the Skyrme constant,
$e = 2\pi$\cite{NuovoCimento}.  The naive, straightforward calculation
of the electromagnetic form factors has shortcomings: the
values of $F_{\pi}$ and $e$ give too small a nucleon size and
the calculated curves do not give the approximate dipole form factor
values.

The generalized Skyrme model under consideration follows the formulation
of Andrianov {\it et al.} \cite{Andrianov2}  This approach uses the
framework of the joint chiral and conformal bosonization of the QCD
Lagrangian, including chiral and scalar dilaton-quarkonium fields.
In such a model the properties of the topological solitons are
dramatically changed in numerical value from those in the original
Skyrme model.  Several authors
have introduced an additional scalar field to the Skyrme model for
different motivational reasons.  For example, Riska and
Schwesinger\cite{Riska88} appear to be the first to
investigate the isospin independent part of the nucleon-nucleon
spin-orbit interaction when a scalar field is added.  A number of papers
studied the effects a scalar $\sigma$ meson would have by introducing
it as a gluon condensate.\cite{Gomm1},\cite{Gomm2},\cite{Jain}, and also
related, \cite{Andrianov2} and \cite{Andrianov3}.  The purely
theoretical and convincing reason is that with the introduction of a
scalar field, the conformal anomaly, one of the distinctive features of
the QCD Lagrangian, is reproduced.  In the SU(2) sector of this
Lagrangian, one can construct an effective theory which reproduces the
conformal anomaly in the framework of the effective Lagrangian method,
introducing a field corresponding to scale invariance.  As shown
in \cite{Lacombe}, \cite{Yabu},and \cite{Mashaal}, it leads to
the necessary strong attraction at intermediate internucleon distances.
In such
an approach the starting point is the fermion integral over quark
fields, in the low energy regime of QCD.  The integral is specified by
the finite mode regularization scheme with a cut-off that also plays the
role of a low energy boundary.  Performing the joint chiral and
conformal bosonization on this integral leads to an effective action for
chiral $U(x)$ and dilaton $\sigma(x)$ fields.  This Lagrangian favors
the linear sigma model in terms of the composite field
$U(x)exp(-\sigma(x))$.  The resulting effective
Lagrangian\cite{Andrianov2}, generalizing the original Skyrme Lagrangian
is
        \begin{eqnarray}
        L_{eff}(U,\sigma) &=& {F_\pi ^2\over 4}exp(-2\sigma )Tr[\partial_\mu
        U\partial ^\mu U^+]+{{N_fF^2_\pi }\over 4}(\partial_\mu\sigma )^2
        \exp(-2\sigma )+\nonumber\\ \cr
        &&+{1\over 128\pi^2}Tr[\partial_\mu U U^+,
        \partial_\nu UU^+]^2-{{C_gN_f}\over 48}(e^{-4\sigma }-1+{4\over
        \varepsilon}(1-e^{-\varepsilon\sigma}))
        \label{c8}\end{eqnarray}
where the pion decay constant is taken as the experimental value,
$F_\pi = 93 MeV$ and $N_f$ is the number of flavors.  The gluon
condensate, according to QCD sum rules, is
$C_g=(300-400 MeV)^4$\cite{Novikov0}.  The first two terms are the
kinetic terms for the chiral and scalar fields and the third term, the
well-known Skyrme term.  The effective potential for the scalar field is
the result of an extrapolation\cite{Andrianov2} of the low energy potential
to high energies by use of a one-loop-approximation to the Gell-Mann Low
QCD $\beta$ - function.  The parameter $\varepsilon$ is determined by
the number of flavors $N_f$ as $\varepsilon={8N_f}/ (33-2N_f)$.

\section{The Nucleon}
\indent
In the baryon sector we choose the chiral field as the spherically
symmetric ansatz of Skyrme and Witten, {$U(\vec x)=exp[-i\vec \tau\vec
n F(r)]$}, where ${\vec n}={\vec r} / |\vec r|$. It is convenient to
introduce a new field, $\rho (x)=exp(-\sigma (x))$.  Then, the mass
functional in dimensionless variables, $x=eF_\pi r$, has the form
$M=M_2+M_4+V$, where
        \begin{eqnarray}
        M_2 &=& 4\pi {F_\pi\over e}\int_{0}^{+\infty}dx[{N_f\over 4}x^2
        (\rho\prime)^2 +\rho ^2({x^2(F')^2\over 2}+sin^2F)]\ ,
        \label{c9}\\ \cr
        M_4 &=& 4\pi{F_\pi\over e}\int_{0}^{+\infty}dx({sin^2F\over {2x^2}} +
        (F\prime ) ^2)sin^2F\ ,
        \label{c10}\\ \cr
        V &=& 4\pi {F_\pi\over e}D_{eff}\int_{0}^{+\infty}dx x^2[\rho^4-1+
        {4\over\varepsilon}\cdot (1-\rho ^\varepsilon )]\ .
        \label{c11}\end{eqnarray}

In the last equations , the same Skyrme parameter value, $e = 2\pi$, is
used.  The contribution of the potential to the mass is determined by
the factor $D_{eff}=C_gN_f/48e^2F_\pi^4$.  The mass functional leads to
a system of equations for the profile functions $F(x)$ and $\rho (x)$,
where a prime is used to denote the derivative with respect to $x$,
        \begin{eqnarray}
        F''[\rho^2x^2+2sin^2F]+2F'x[x\rho\rho '+\rho ^2]+(F')^2\cdot
        sin(2F)-\nonumber\\ \cr
        -\rho ^2\cdot sin(2F)-sin(2F)\cdot sin^2F/x^2 =0\ ,
        \label{c12}\end{eqnarray}
        \begin{eqnarray}
        {N_f\over 2}x[x\rho ''+2\rho ']-2\rho [{x^2(F')^2\over 2}+sin^2F] -
        \nonumber\\ \cr
        - 4D_{eff} \cdot [\rho ^3-\rho^{\varepsilon -1}] x^2= 0\ ,
        \label{c13}\end{eqnarray}
At small distances, $F = \pi N-\alpha x$ and $\rho = \rho (0)+\beta
x^2$, with $\rho (0) \neq 0$.  For large $x$, these functions behave as
$F(x)\sim a/x^2$, and $\rho(x)\sim 1-b/x^6 + \dots$.

\begin{figure}[htb]
 \centerline{ \includegraphics[width=125mm]{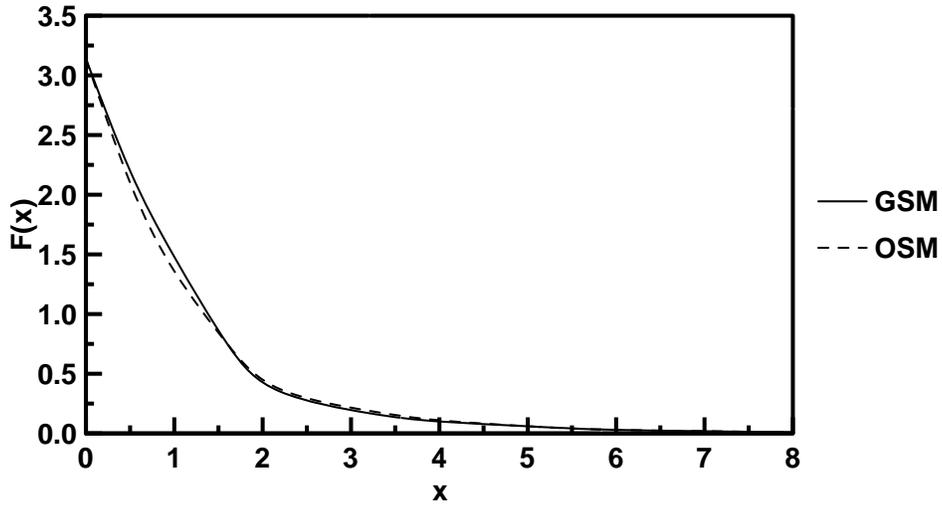}}
 \caption{Chiral angle $F(x)$ in GSM and OSM for $C_g=(300~Mev)^4$ and$N_f=2$.}
 \label{chiralangle}
\end{figure}

\begin{figure}[htb]
 \centerline{ \includegraphics[width=125mm]{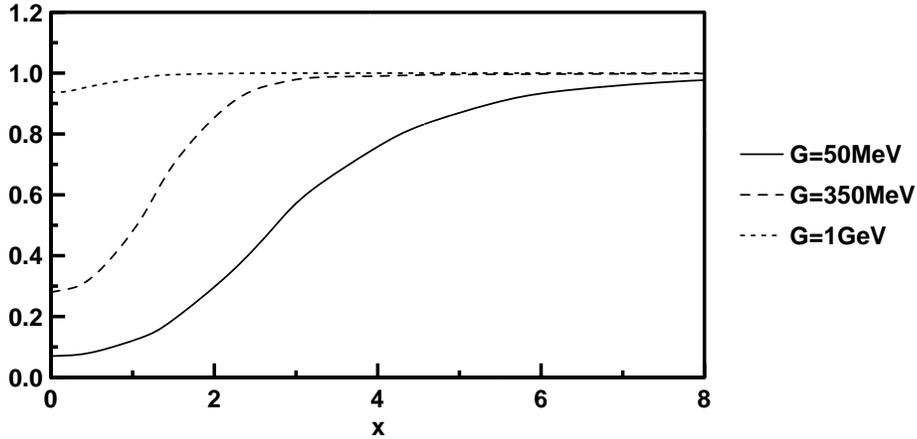}}
 \caption{Scalar meson shape function $rho(x)$ for $G=C_g^{1/4}=50~Mev$,  $350~Mev$ and $1~GeV$.}
 \label{rhofunction}
\end{figure}

\begin{figure}[htb]
 \centerline{\includegraphics[width=125mm]{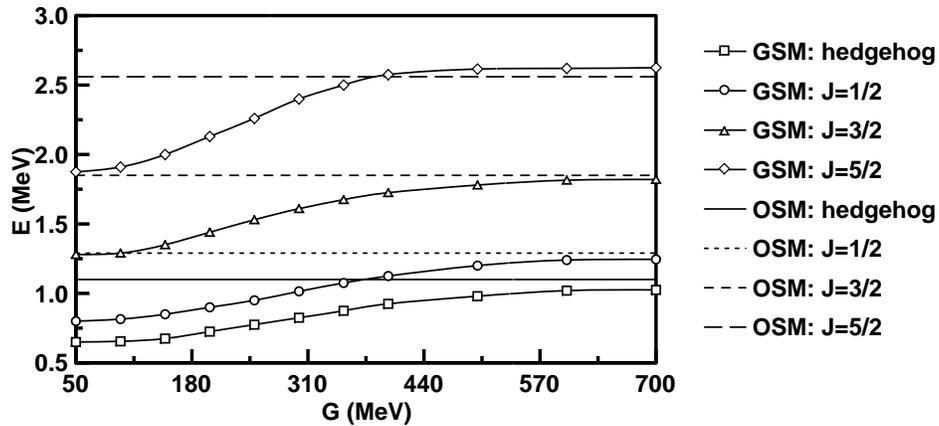}}
 \caption{Mass spectra of the ground and excited states in GSM (solid line) and OSM (dashed line).}
 \label{massfromG}
\end{figure}

According to the virial theorem,\cite{Nikolaev109} the contributions of
the individual terms of the mass functional to the energy of
the system must obey the condition,
\begin{eqnarray} M_4
- M_2 -3V = 0 \ , \label{c13a}\end{eqnarray}
which can be used to control the accuracy of the numerical solution of
the system.  There are nontrivial equations between the numbers
 $\alpha$ and $\beta$, $a$ and $b$,
 \begin{eqnarray} b &=&
        {1\over 2} a^2/D_{eff} \ , \label{c13b}\\ \cr \beta &=&
        \left[\rho(0)\alpha^2+{4\over 3}\left(\rho^3(0)-1\right)
        D_{eff}\right]/N_f\ .
        \label{c13c}\end{eqnarray}
The choice of boundary conditions ensures a finiteness of the mass functional for a
given value of the topological charge $B = N$.
Performing canonical quantization of the rotational degrees of freedom
with the collective variable method,\cite{Witten} one obtains for the
nucleon mass,
        \begin{eqnarray}
        M_B= M+S(S+1)/(2I)\ ,
        \label{c14}\end{eqnarray}
where the moment of inertia is
        \begin{eqnarray}
        I={8\pi\over 3}(F_\pi e)^{-3}\int_{0}^{\infty}dx\ sin^2[\rho^2x^2 +
        (F')^2 x^2+sin^2F]\ .
        \label{c15}\end{eqnarray}

\noindent
Some numerical results are presented in Table 1,
where the soliton mean square radius of the corresponding baryon
$<r^2_{IS}>$ density distribution is given
 \begin{eqnarray}
        <r^2_B>^{1/2}={1\over F_\pi e}\biggl\{-{2\over\pi}
\int_{0}^{\infty}dx
        x^2F' sin^2F\biggr\}^{1/2}\ .
        \label{c16}\end{eqnarray}

\begin{table}[htb]
\begin{center}
\begin{minipage}{3.5cm}
\begin{tabular}{|c|c|}
& Present work \\
\hline
 $M$        & 839 $MeV$ \\
 $<r^2>_p $ & 0.67 $Fm^2$\\
 $<r^2>_n$  & -0.14 $Fm^2$\\
 $M_B$      & 1026 $MeV$\\
\end{tabular}
\end{minipage}
\end{center}
\caption{Static properties ($N_f = 2$) in the generalized Skyrme
model with $F_\pi =93 MeV,\ e=2\pi ,\ C_g=(300 MeV)^4$. }
\end{table}

A discussion of partial restoration of chiral symmetry in this model is
given in Ref.(\cite{NuovoCimento}. The restoration appears as a large
deviation of $\rho(0)$ from its asymptotic value of $\rho(0)=1$.  The
dependence of the mass spectra on the gluon condensate in the
generalized Skyrme model was also discussed in \cite{NuovoCimento}.


\section{Form factors of charge distributions}
\indent
The nucleon electric and magnetic form factors, $G_E(q^2) $and
$G_M(q^2)$ can be calculated from the electromagnetic currents, in
the Breit frame where the photon does not transfer energy.
\begin{eqnarray}
<N_f({\vec q\over 2})| J_0(0) |N_i(-{\vec q\over 2})> &=& G_E(\vec
q^2) \xi_f^+ \xi_i\ ,\nonumber\\
<N_f({\vec q\over 2})| \vec J(0) |N_i(-{\vec q\over 2})> &=&
{G_M(\vec q^2)\over 2M_N} \xi_f^+ i\vec\sigma\otimes\vec q\xi_i\ .
\label{F2}
\end{eqnarray}
\noindent
Here, $|N(\vec p)>$ is the nucleon state with momentum $\vec p$,
$\xi_i$, $\xi_f$ and two component Pauli spinors, and $\vec q \equiv $
momentum transfer.

The isoscalar (S) and isovector (V) nucleon form factors are related to
those for the proton and neutron by
\begin{eqnarray}
G^{{\rm p,\ n}}_{E,M}=G^{S}_{E,M} \pm G^{V}_{E,M}
\label{SVfactors}
\end{eqnarray}

These form factors are normalized to the respective charge and magnetic
moments by
\begin{eqnarray}
&&G^{\rm p}_{E}(0)=1 \qquad G^{\rm n}_{E}(0)=0  \cr
&&G^{\rm p}_{M}(0)\equiv \mu_p=2.79 \qquad G^{\rm n}_{M}= \mu_n=-1.91\ .
\label{Norm}
\end{eqnarray}
We remaked above on the smallness of the nucleon size as determined
by the baryon charge density distribution in the model with a
dilaton-quarkonium field.

Vector meson dominance means that the isoscalar photon sees
$\omega$ meson ${\it structure}$, but not the isoscalar baryon density
$B_0(r)$.

According to vector meson dominance, the isoscalar current is
proportional to the $\omega_{\mu}$-field,
\begin{eqnarray}
J^{\mu}_{I=0}=-\frac{m^2_{\omega}}{3g}\omega_{\mu}(r)
\label{ICurrent}
\end{eqnarray}
and the corresponding charge form factor,
\begin{eqnarray}
G^{S}_{E}(\vec q^2)=-\frac{m^2_{\omega}}{3g}\int d^3r\exp{i\vec q\vec r}\omega(r)\ .
\label{IFactor}
\end{eqnarray}

The static $\omega (r)$ obeys the equation,
\begin{eqnarray}
(\nabla^2-m^2_{\omega})\omega (r)=\frac{3g}{2}B(r)=-\frac{3gF'(r)}{4\pi r^2}sin^2F(r)\ .
\label{EqForOmega}
\end{eqnarray}
From this equation, we obtain,
\begin{eqnarray}
G^{S}_{E}(\vec q^2)=-\frac{1}{2}\frac{m^2_{\omega}}{m^2_{\omega}+
\vec q^2}4\pi\int drr^2B_0(r)j_0(qr)\ .
\label{SFormFact}
\end{eqnarray}
Therefore, the effecive isoscalar nucleon density is equal baryon charge
density $B_0(r)$ times the $\omega$-meson propagator.

The isovector electromagnetic formfactor has analogous structure,
\begin{eqnarray}
G^{V}_{E}(\vec q^2)=-\frac{1}{2}\frac{m^2_{\rho}}{m^2_{\rho}+\vec q^2}F^V_E(q^2)\ ,
\label{VFormFact}
\end{eqnarray}

In writing the propagators separately, as a factor
multiplied into  $F^V_E$, $\omega$ and $\rho$, themselves have no
substructure or internal dynamics; the corresponding Skyrmion densities
are considered as the sources of these $\omega$ and $\rho$ fields.
Explicit considerations of the role of vector mesons in the
electromagnetic form factors in the $\sigma$ model has been given by
Holzwarth\cite{HOLZ} and quantum corrections to the relevant baryon
properties in the chiral soliton models has been calculated.\cite{MEIER}

\begin{figure}[htb]
 \centerline{
 \includegraphics[width=125mm]{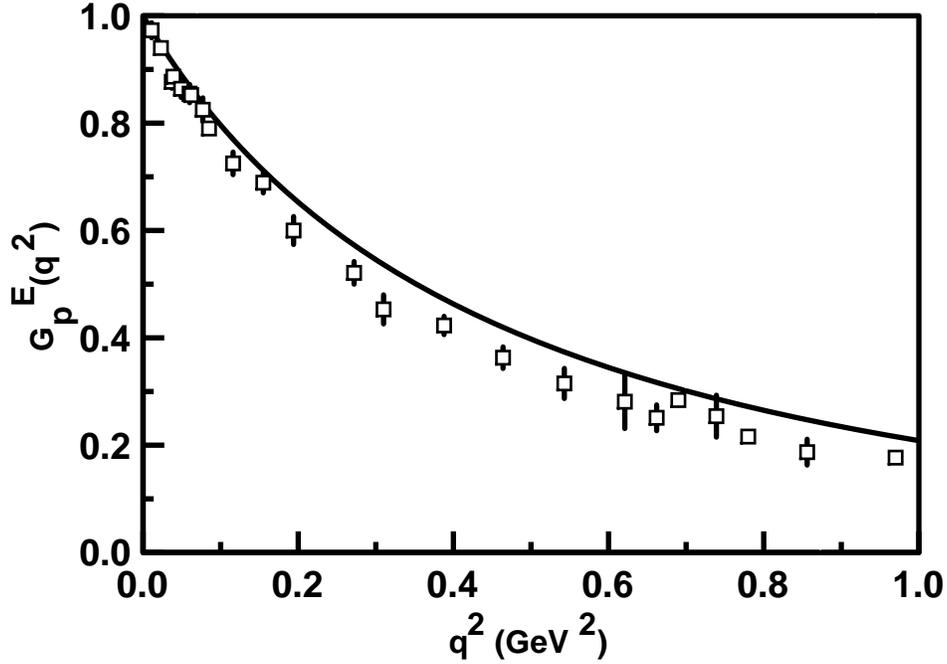}}
 \caption{Proton electric form factor as a function of $q^2$ in $GeV^2$
calculated for $F_\pi = 93$, $e=2\pi$, $N_f=2$, $C_g=(300 MeV)^4$,
and $m_\pi=139$. The experimental data shown come from
Ref.~\protect\cite{Bartel}.} \label{feprot}
\end{figure}

\begin{figure}[htb]
 \centerline{
 \includegraphics[width=125mm]{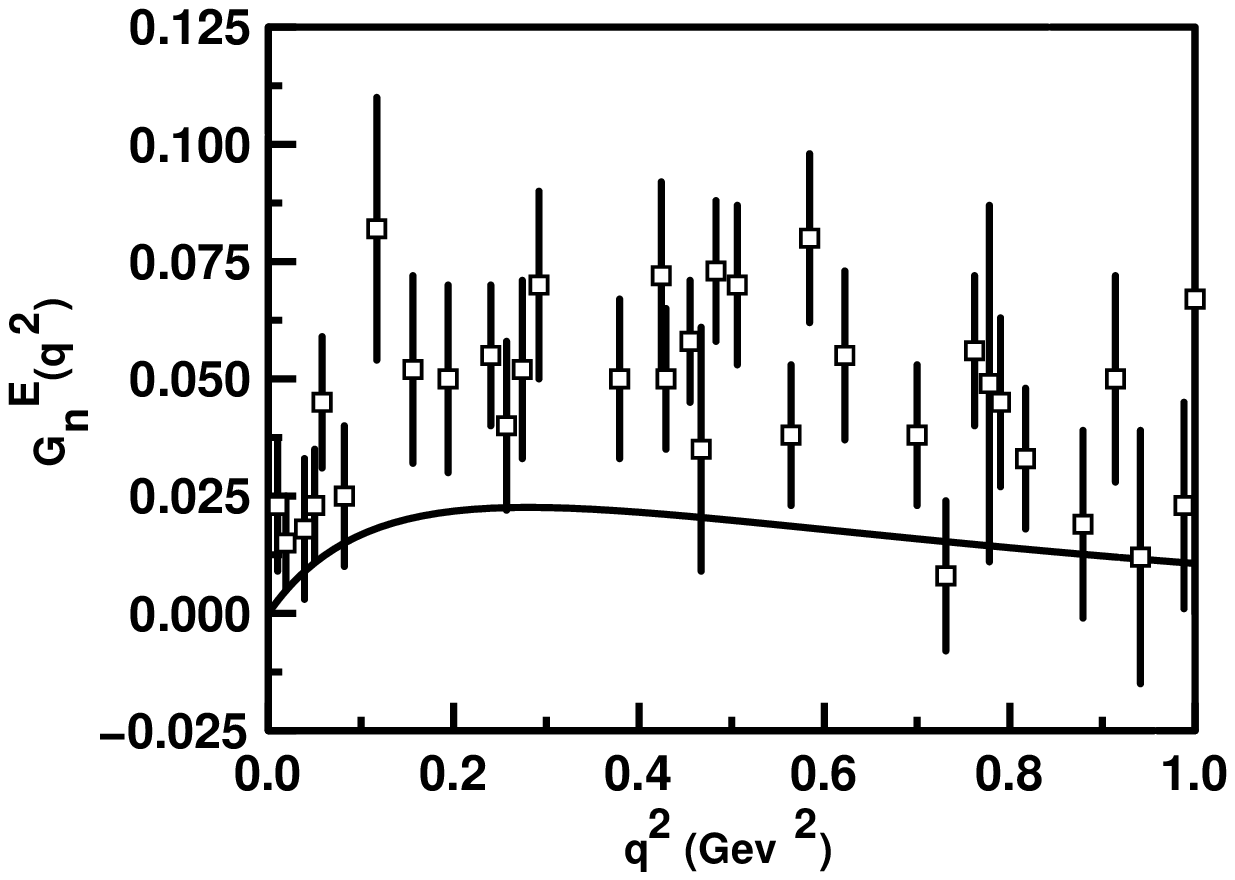}}
 \caption{Neutron electric form factor as a function of $q^2$ in $GeV^2$
calculated for $F_\pi = 93$, $e=2\pi$, $N_f=2$, $C_g=(300 MeV)^4$,
and $m_\pi=139$. The experimental data shown come from
Ref.\protect\cite{Bartel}.\label{feneut}}
\end{figure}

\begin{figure}[htb]
 \centerline{
 \includegraphics[width=125mm]{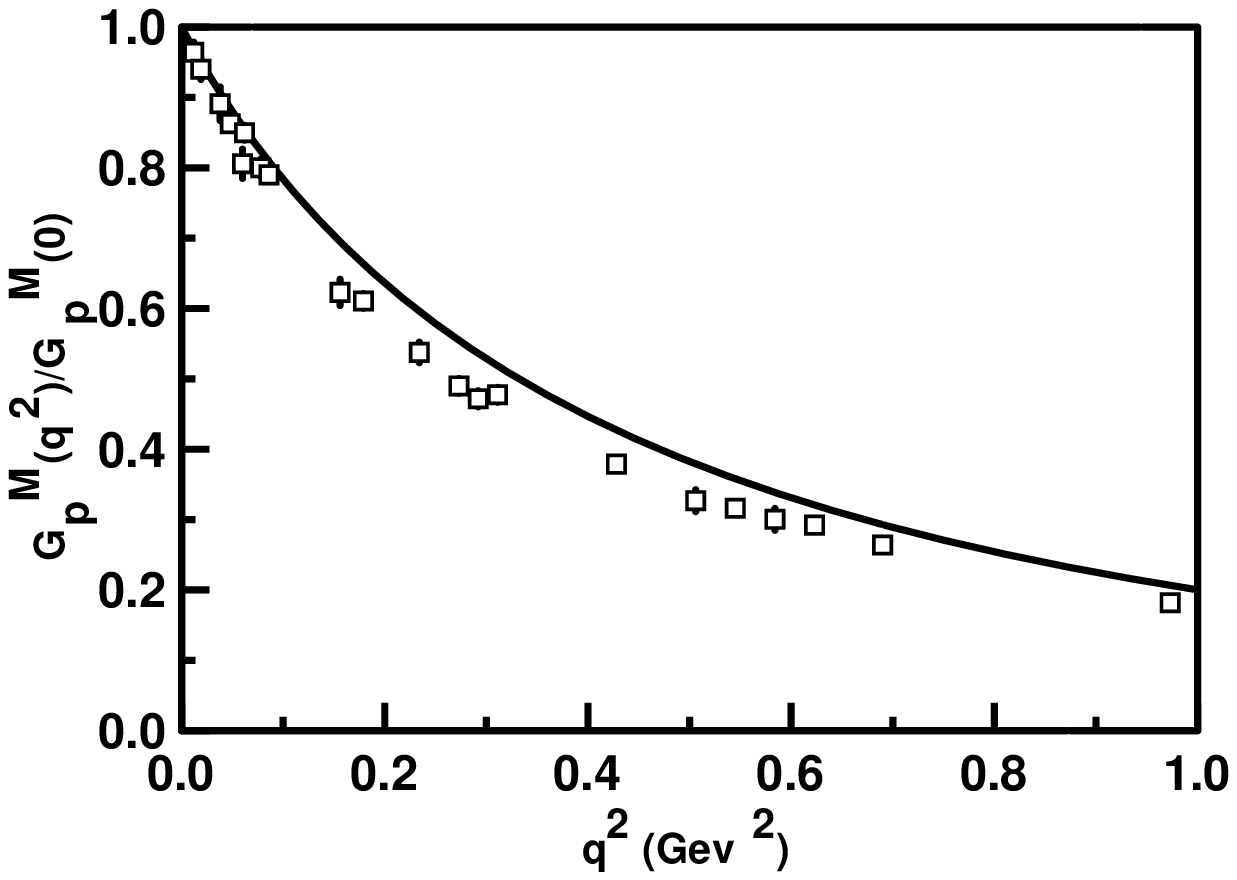}}
 \caption{Proton magnetic form factor as a function of $q^2$ in $GeV^2$
calculated for $F_\pi = 93$, $e=2\pi$, $N_f=2$, $C_g=(300 MeV)^4$,
and $m_\pi=139$. The experimental data shown come from
Ref.\protect\cite{Bartel}.\label{fmprot}}
\end{figure}

\begin{figure}[htb]
 \centerline{
 \includegraphics[width=125mm]{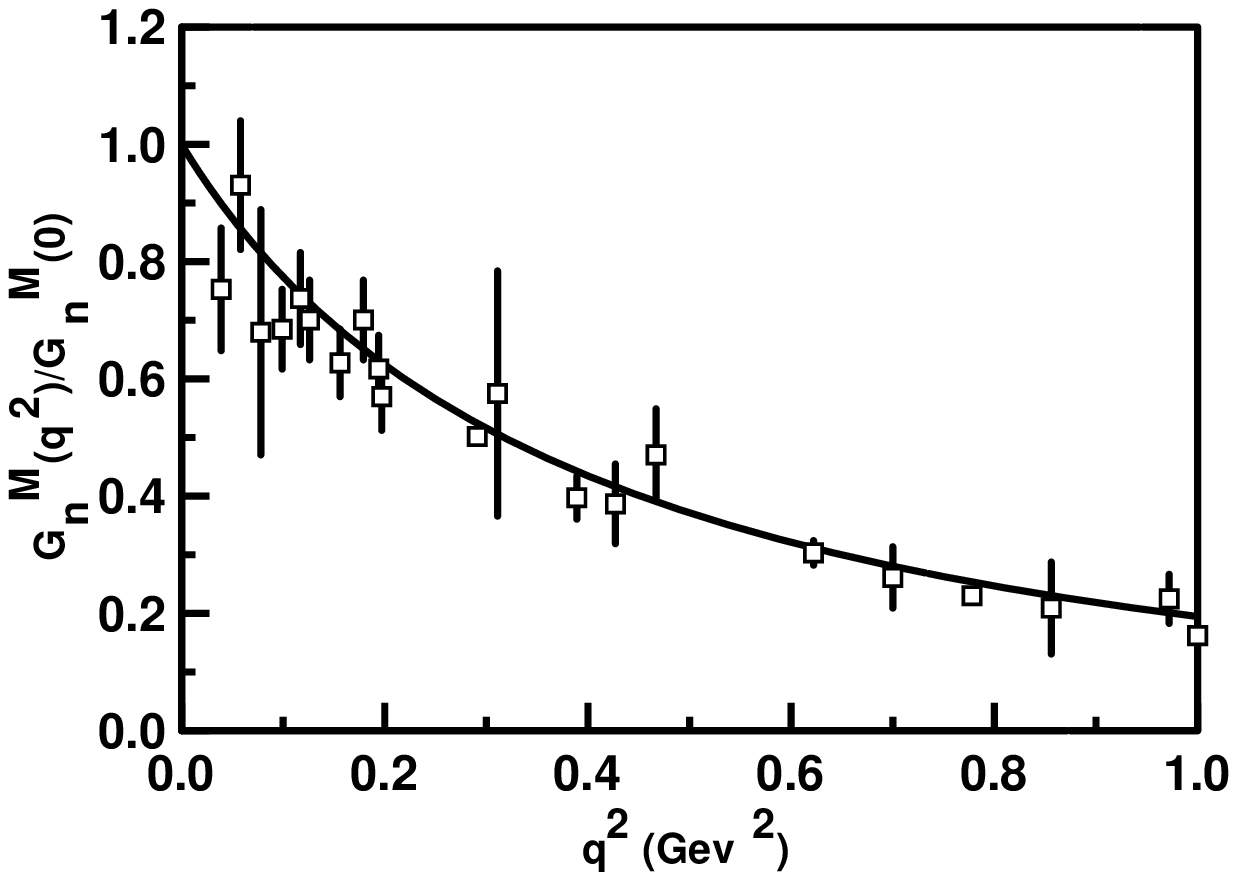}}
 \caption{Neutron magnetic form factor as a function of $q^2$ in $GeV^2$
calculated for $F_\pi = 93$, $e=2\pi$, $N_f=2$, $C_g=(300 MeV)^4$,
and $m_\pi=139$. The experimental data shown come from
Ref.\protect\cite{Bartel}.} \label{fmneut}
\end{figure}

The results of the present calculations are given in Figures \ref{feprot} to \ref{fmneut}.
The isoscalar part of the Skyrmion electric charge coincides with the
baryon density distribution, and for the isovector density from the
Skyrmion model one obtains,
        \begin{eqnarray}
        \rho^{V}(x) = sin^2F(x)\left[x^2\rho^2(x) +
        \left(F^\prime(x)\right)^2x^2 + sin^2F(x)\right]\ .
        \label{c17}\end{eqnarray}
To take chiral symmetry breaking into account, we must add the
pion mass term,
        \begin{eqnarray}
        {\cal L}_\pi = {1\over 4} m_\pi^2 F_\pi^2 e^{-3\sigma}
        Tr\left[U + U^+ - {3\over 2} e^{-\sigma}\right]\ ,
        \label{c19}\end{eqnarray} to our Skyrme model Lagrangian.
The theoretical predictions for the proton electric, neutron
electric, proton magnetic and neutron magnetic form factors,
compared with data are shown in Figures \ref{feprot},
\ref{feneut}, \ref{fmprot} and \ref{fmneut} respectively.
Corresponding values of the proton and neutron mean square radius
of the electric charge distribution are 0.78 $Fm^2$ and - 0.19
$Fm^2$.

\indent
We have presented our calculations on the nucleon electromagnetic form
factors in the framework of the generalized Skyrme model with dilaton
quarkonium field. The first calculation~\cite{NuovoCimento} in
such a model yielded large deviations of the calculated form factors
from the dipole approximation formula.  We would like to point out here that
 we use the
empirical value of the pion decay constant and the theoretical value for
the Skyrme term constant in the vector meson dominance approach to
obtain a good description of the form factor data in the finite
range of momentum transfer in the measurements. The vector mesons
are included only as elements of the hadron substructure of the photon
and are not considered as components of the structure in the soliton
self-dynamics.  Implicit in the approach, though not explicitly
proposed, is the possibility of having the role of vector mesons given
by higher derivative terms in the effective Lagrangian for soliton
dynamics~\cite{Rajat}.  For example, keeping terms to four
orders in the expansion of the effective Lagrangian would lead to
a $\rho$ meson-like term and the sixth order terms would give
 $\omega$-like terms which are important in the calculations
of the form factors at the larger momentum transfers.

Lagangian  including $\omega$--field becomes
\bea \label{skydil}
 {\cal L}  =  \L_{eff}(U,\sigma)
- V_{\s}- {1\over 4} {G_{\mu\nu}G^{\mu\nu}}
+ \half~e^{2\s} m_\o^2~~\o_\mu^2 -g_V~~ \o_{\mu} B^{\mu}
\eea
Here
\bea \label{rho}
G_{\mu\nu} = \dd_{\mu}  \o_{\nu} - \dd_{\nu}  \o_{\mu}
\eea

To discusse skyrmion properties in nuclear interiour one can use
the  ans\"atze for the scalar and vector fields
\bea \label{ansatz}
U(\bf r) & = & \exp[i\bftau\cdot\hat{\bf r} F(r)], \nono
\o^{\mu}(\bf R) & = & \delta_{{\mu}0}~\o(R)\nono
\eea
where ${\bf R}$ measures the distance from the center of the nucleus at rest,
and ${\bf r}$ is the coordinate from the center of the skyrmion.

To introduce the mean field -like approximation to for the meson fields
one can use
\bea \label{additivity}
\s_{B=N} & = & \s_1 + \s_2 + \cdots+ \s_N, \nono
& = &\s_0 + \delta\s_1 + \delta\s_2 +\cdots+ \delta\s_N, \\
\o_{B=N} & = & \o_1 + \o_2 + \cdots+ \o_N, \nono
& = &\o_0 + \delta\o_1 + \delta\o_2 +\cdots+ \delta\o_N,
\eea
where $\s_0 , \o_0$ are the mean field constant values of the
scalar and the $\o$ and $\delta\s, \delta\o$ represent the
fluctuations.

The topological baryon density
\be \label{Bmu}
B^\mu = \frac{\epsilon^{\mu\alpha\beta\gamma}}{24\pi^2}
\tr \left[\left(U^\da\dd_\alpha U\right)
\left(U^\da\dd_\beta U\right)
\left(U^\da\dd_\gamma U\right)\right],
\ee
with the the product ansatz will give us
\bea \label{b_1}
B_0 & = & b_1 + b_2 + \cdots + b_N,\nono
\eea

Skyrmion mass becomes depending on position R:
\bea \label {mass}
 M(R) & = & 4\pi\int_0^\infty r^2\,dr M(r)\nono
M(r) &= &\e^{2 \s}\frac{F_\pi^2}{8}
\left[(F')^2 + 2 \frac{\ssF}{r^2}\right] + \frac{1}{2 e^2}
\frac{\ssF}{r^2} \bigg[\frac{\ssF}{r^2}+2
(F')^2\bigg]
\eea

A key element of the nucleon interaction in the nucleus is the
spin-orbit potential if we want to construct shell model -like theory.
There is very important  source of spin-orbit interaction\cite{Kalbermann}
\bea\label{uso2}
W_{s.o.} = \frac{-{\bf S} \cdot {\bf L}}{2~M_0~\lambda(R)}~\o_1(R)\ .
\eea
It is due to the
transformation of the fields to a rotating frame, essentially the
coupling of the baryon current to the $\o$ field in a rotating
nucleus analogous to the isoscalar coupling to the photon\cite{Kalbermann}.

In the Dirac type of
Walecka models \ci{ser1}, the spin-orbit interaction arises from
the coupling of the lower components of the Dirac wave function.
Here, it arises from the interaction of the rigid rotation of the nucleon
with the flow of the mean fields, which is quite a different mechanism.

 So  one could decide that we are on the way to construct nuclear shell model
on Skyrme model background.
 But using the same arguments and R- dependence of sigma-field we can
obtain that the nucleon swells inside the nucleus to almost twice its free
size\cite{Kalbermann}. This is a big trouble in our execises.
The nucleons no longer act as free enteties and the solitons overlap.

It is the only  way to avoid the problem - to consider the interiour of nuclei as
one soliton with complicated structure.
Let consider original Skyrme model \cite{Skyrme1} without dilaton and omega -fields

     Here we follow  the paper \cite{Nikolaev5}
(see also \cite{Nikolaev7}).
 For a variational treatment we use the chiral field $U$
\begin{eqnarray}
U(\vec r ) = \cos F(r) + i(\vec {\tau} \cdot \vec N )\ \sin F(r) .
\label{e1}
\end{eqnarray}
with the following assumption about the configuration of the isotopic vector
field $\vec N$ :
\be
\vec N = \{ \cos(\Phi (\phi ,\theta ))  \sin(T(\theta )),
\ \sin(\Phi (\phi ,\theta ))  \sin(T(\theta )),\ \cos(T(\theta )) \} .
\label{e2}
\ee
Here $\Phi (\phi,\theta),\ T(\theta)$ are some
arbitrary functions of angles $(\theta ,\phi)$ of the vector $\vec r$
in the spherical coordinate system.

\begin{figure}[htb]
 \centerline{
 \includegraphics[width=125mm]{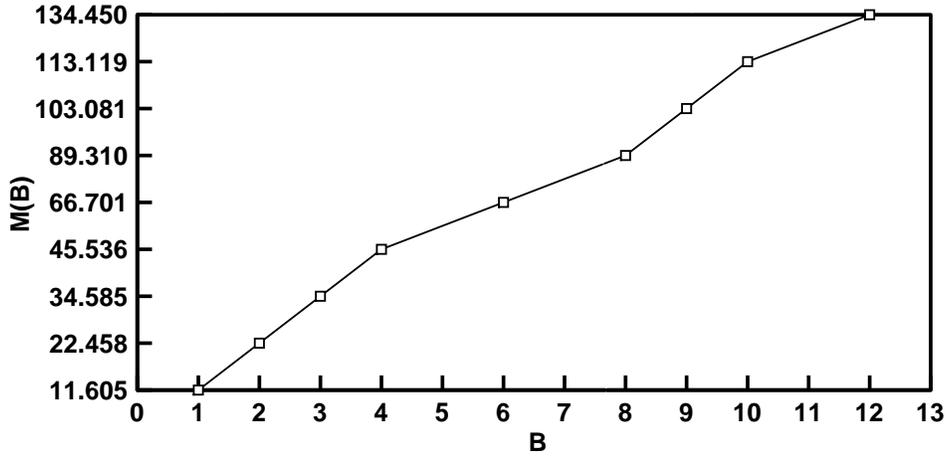}}
 \caption{Dependence of the classical masses $M$ in dimensionless variables
 $\pi{F_\pi\over e}$ on the baryon charge $B$. \label{linearmassofbaryons}}
\end{figure}

Let us consider the Lagrangian density ${\cal L}$ for the stationary
solution:
\be
{{\cal L}} = {F_\pi^2\over 16}  {\rm Tr}(L_\mu L^\mu)+{1\over 32e^2}
{\rm Tr}\Bigl[L_\mu,L_\nu\Bigr ]^2 .\label{e3}
\ee
Here $L_\mu= U^+{\partial}_\mu U$ are the left currents.

Variation of the functional $L=\int {\cal L}d\vec r$ with respect to
$\Phi$ leads to an equation which has a solution of the type
\be
\Phi(\theta ,\phi) = k(\theta) \phi+c(\theta)
\ee
with a constraint
\be
{\partial\over\partial\theta}\left[\sin^2T(\theta)  \sin\theta
{\partial \Phi(\theta,\phi)\over\partial\theta}\right] = 0. \label{e8}
\ee
It is easily seen from eq. (\ref{e8}) (see also \cite{Nikolaev5})
that functions $k(\theta)$ and $c(\theta)$ may be piecewise constant
functions.

Moreover, $k^{(m)}$ must be integer in any region
$\theta_m\leq\theta\leq\theta_{m+1}$,
where $\theta_m$, $\theta_{m+1}$ are successive points of discontinuity
of $\partial \Phi_i(\theta ,\phi) / \partial \theta$.
The positions of these are the points determined by the condition
\be
T(\theta_m) = m \pi\ ,\ \ \ T(\pi) = n \pi \label{e9}
\ee
with integer $m$, as follows from Eq.(\ref{e8}).

The soliton mass is given by a functional which can be represented as a
sum of contributions from different $\theta$ - regions.
The functions $F (x)$ and $T (\theta)$ have to obey the equations
derived in \cite{Nikolaev5}, in each $\theta$ - region with given
number $k^{(m)}$.

We obtain  an almost linear dependence of the classical masses $M$ on the
baryon charge $B$.  The results of the calculations are given in Figure \ref{linearmassofbaryons}.

\section{Conclusions}
\indent
The one particle ( B=1) degrees of freedom
are concentrated around  the surface of the nuclei and the soliton with  non-trivial structure ( $B >$1) is located
at the center region.


\begin{thebibliography}{18}
\bibitem{NuovoCimento}Nikolaev V.A, Tkachev O.G., Novozhilov V.Yu,
                  IL NUOVO CIMENTO 1994. V. 107A(12). P.2637.
\bibitem{Andrianov2}  Andrianov A.A., Andrianov V.A., Novozhilov Yu.V.,
                      Novozhilov V.Yu.
                      Phys. Lett. 1987. V. B186(3,4). P.401.
\bibitem{Riska88}     Riska D.O. and Schwesinger B.
                      Phys. Lett. 1989. V. B229. P.339.
\bibitem{Gomm1}       Gomm M., Jain P., Jonson R. and Schehter J.
                      Phys. Rev. 1986. V. D33(3). P.801.
\bibitem{Gomm2}       Gomm M., Jain P., Jonson R. and Schehter J.
                      Phys. Rev. 1986. V. D33(11). P.3476.
\bibitem{Jain}        Jain P., Jonson R. and Schechter J.
                      Phys. Rev. 1987, V. D35(7). P.2230.
\bibitem{Andrianov3}  Andrianov A.A., Novozhilov Yu.V.
              Phys. Lett. 1988. V. B202. P.580.
\bibitem{Lacombe}     Lacombe M. Loiseau B., Vinh Mau, Cottingham W.N.
                      IPN Orsay preprint 87-28. 1987.
\bibitem{Yabu}        Yabu H., Schwesinger B., Holzwarth G.
                      Phys. Lett. 1989. V. B224. P.25.
\bibitem{Mashaal}     Mashaal M., Phom  T.N., Truong T.N.
                      Phys. Rev. 1986. V. D34. P.3484.
\bibitem{Novikov0}    Novikov V.A., Shifman  M.A.,  Vainstein  A.,
                      Voloshin  M.B., Zakharov V.I.
                      Nucl. Phys. 1984. V. B237. P.525.
\bibitem{Nikolaev109} Nikolaev V.A. Proceedings of the IX ISHEPP. 1988.
              Dubna. V. 1. P.51.
\bibitem{Witten}      Adkins G.S., Nappi C.R. and Witten E.
              Nucl. Phys. 1983. V. B228(3). P.552.
\bibitem{HOLZ}        Holzwarth G.
              Contribution to the Sixth International Symposium on
              Meson-Nucleon Physics and the Structure of the Nucleon.
              Blaubeuren/Tuebingen. Germany. 10-14 July. 1995.
\bibitem{MEIER}       Meier F., Walliser H.
              Physics Reports. 1997. V. 289. P.448.
\bibitem{Bartel}      Bartel W. et al. Nucl. Phys. 1973. V. B58. P.429.
              Kirk P.N. et al. Phys. Rev. 1973. V. D8. P.63.
              Hanson K. et al. Phys. Rev. 1973. V. D8. P.753.
              Rock S. et al. Phys. Rev. Lett. 1982. V. 49. P.1139.
              Arnold R.G. et al. Phys. Rev. Lett. 1986. V. 57. 174.
              Bosted P.E. et al. Phys. Rev. 1990. V. C42. P.38.
\bibitem{Rajat}       Rajat K. Bhaduri Lecture Notes and Supplements in
              Physics:  Models of Nucleon. Addison-Wesley
              Publishing Company. Inc. 1988. P.283.
\bibitem{Kalbermann} G. Kalbermann
HEP preprint Nucl-th/9808059,
Nucl.Phys. A612 (1997) 359-374
\bi{ser1} B. D. Serot and J. D. Walecka, Adv. Nucl. Phys. {\bf 16} (1986) 1.
\bi{riska} D. O. Riska and B. Schwesinger, Phys. Lett. {\bf B229} (1989) 339.
\bibitem{Skyrme1} T.H.R.Skyrme, Nucl.Phys. 31 (1962) 556
\bibitem{Nikolaev5} Nikolaev V.A., Tkachev O.G.
        Sov.J.Part.Nucl. {\bf 21} N6 (1990) 643.
\bibitem{Nikolaev7} R.M.Nikolaeva, V.A.Nikolaev, O.G.Tkachev,
        Jour. of Nucl. Phys. 56 (1993) 173.\\
        R.M.Nikolaeva, V.A.Nikolaev, O.G.Tkachev,
        J.Phys.G: Nucl.Phys., 18 (1992) 1149.
\end{thebibliography}
\end{document}